%% file: main.tex
\documentclass[sigconf]{_acm/acmart}

\usepackage{bigstrut}
\usepackage{booktabs}
\usepackage{multirow}
\usepackage{microtype}
\usepackage{graphicx}
\usepackage{subfig}
\usepackage{bbding}
\usepackage{overpic}
\usepackage{balance}
\usepackage{amsthm}
\usepackage{array}
\usepackage{clrscode}
\usepackage{multicol}
\usepackage{float}
\usepackage{newfloat}
\usepackage{color}
\usepackage[table]{xcolor}
\usepackage{amsopn}
\usepackage{mathrsfs}
\usepackage{mathtools}
\usepackage{amsmath}
\usepackage{arydshln}
\usepackage{blkarray}
\usepackage{enumerate}
\usepackage{courier}
\usepackage{rotating}
\usepackage{bm}
\usepackage{wrapfig}
\usepackage{ragged2e}
\usepackage[misc]{ifsym}
\usepackage{threeparttable}
\usepackage{makecell}
\usepackage{adjustbox} 
\usepackage{enumitem}
\usepackage{url}
\usepackage{xspace}
\usepackage{comment}
\usepackage{changepage}
\usepackage{algorithm}
\usepackage{algpseudocode} 
\usepackage{algorithmicx}

\usepackage{amssymb}
\usepackage{utfsym}
\usepackage[ruled, vlined, nofillcomment, linesnumbered, algo2e]{algorithm2e}
\usepackage[normalem]{ulem}
\usepackage{hyperref}

\useunder{\uline}{\ul}{}

\newcommand{\ie}{\emph{i.e.,}\xspace}
\newcommand{\eg}{\emph{e.g.,}\xspace}

\newcommand{\paratitle}[1]{\vspace{1.5ex}\noindent\textbf{#1}}
\newcommand{\wrt}{w.r.t.\xspace}

\newcommand{\ignore}[1]{}

\AtBeginDocument{%
  \providecommand\BibTeX{{%
    \normalfont B\kern-0.5em{\scshape i\kern-0.25em b}\kern-0.8em\TeX}}}

\copyrightyear{2026}
\acmYear{2026}
\setcopyright{cc}
\setcctype{by}
\acmConference[KDD '26]{Proceedings of the 32nd ACM SIGKDD Conference on Knowledge Discovery and Data Mining V.2}{August 09--13, 2026}{Jeju Island, Republic of Korea}
\acmBooktitle{Proceedings of the 32nd ACM SIGKDD Conference on Knowledge Discovery and Data Mining V.2 (KDD '26), August 09--13, 2026, Jeju Island, Republic of Korea}
\acmDOI{10.1145/3770855.3818179}
\acmISBN{979-8-4007-2259-2/2026/08}

\begin{document}

\title{Collaborative Memory Augmentation for Generative Recommendation}

\author{Enze Liu$^\dagger$}
\orcid{0009-0007-8344-4780}
\affiliation{%
    \department{Gaoling School of Artificial Intelligence}
    \institution{Renmin University of China}
    \city{Beijing}
    \country{China}
}
\email{enzeliu@ruc.edu.cn}

\author{Zhen Tian}
\orcid{0000-0001-5569-2591}
\affiliation{%
    \institution{ByteDance}
    \city{Beijing}
    \country{China}
}
\email{chenyuwuxinn@gmail.com}

\author{Wayne Xin Zhao\textsuperscript{\Letter}}
\orcid{0000-0002-8333-6196}
\affiliation{
    \department{Gaoling School of Artificial Intelligence}
    \institution{Renmin University of China}
    \city{Beijing}
    \country{China}
}
\email{batmanfly@gmail.com}

\thanks{$^\dagger$ Work done during an internship at ByteDance.}
\thanks{\Letter \ Corresponding author.}
\renewcommand{\shortauthors}{Enze Liu, Zhen Tian, \& Wayne Xin Zhao}

\begin{abstract}

Generative Recommendation (GR) has exhibited great potential by modeling item transitions as a sequence-to-sequence task. Despite the success of GR, existing frameworks primarily focus on modeling individual user sequences within a constrained internal parametric space, failing to explicitly leverage cross-user collaborative signals. To address this issue, we propose \textbf{OMEGA}, a c\underline{O}llaborative \underline{ME}mory augmentation framework for \underline{G}enerative recommend\underline{A}tion. OMEGA bridges the gap between implicit parametric knowledge and explicit collaborative signals. We first introduce a latent context compression method that utilizes learnable query tokens to distill sequential user behavior into compact representations, significantly reducing storage overhead. These compressed representations are aggregated into a collaborative memory bank, serving as an explicit repository of global behavioral patterns. To ensure precise knowledge acquisition, we design a lightweight and target-aware retrieval mechanism that identifies pertinent memories by considering both sequence-level and target-level similarities. Furthermore, a context-aware integration module, equipped with a gated cross-attention mechanism, is employed to adaptively fuse the retrieved collaborative memories with the local user context while mitigating the interference of noisy patterns. Empirical evaluations on multiple real-world datasets demonstrate that OMEGA significantly outperforms existing advanced GR models, validating the potential of external memory as a complement to the generative paradigm.

\end{abstract}



\maketitle

\input{sections/1-introduction.tex}

\input{sections/2-methodology.tex}
\input{sections/3-experiments.tex}

\input{sections/4-related-work.tex}
\input{sections/5-conclusion.tex}

\begin{acks}
This work was supported by the National Natural Science Foundation of China No.~92470205 and Beijing Major Science and Technology Project No.~Z251100008425002.
\end{acks}

\bibliographystyle{_acm/ACM-Reference-Format}
\balance
\bibliography{bibliography}

\appendix
\input{sections/6-appendix.tex}

\end{document}

%% file: sections/1-introduction.tex
\section{Introduction}
\label{sec:introduction}

Recently, Generative Recommendation (GR) has received significant attention within both academic and industrial communities~\cite{onerec,tiger,gpr,lc-rec,etegrec,genrank}. The key difference in GR resides in the mechanisms of item tokenization and generative retrieval.
In contrast to traditional ID-based methodologies, GR models represent each item as a sequence of discrete tokens within a shared vocabulary space, also referred to as semantic IDs.
Subsequently, a generative model is trained to generate predictions in an autoregressive manner.
This design allows GR models to benefit from the powerful sequential modeling capability of the LLM architecture and exhibit superior scaling behavior~\cite{trm}.
Furthermore, it demonstrates the potential to replace traditional cascaded multi-stage pipelines with a unified end-to-end model~\cite{onerec}.

Despite their promise, existing GR approaches typically operate in an isolated manner, modeling the probability of the next item solely based on the interaction history of an individual user. These methods implicitly rely on static internal parameters to encapsulate the vast and complex collaborative patterns across the entire user base.
However, compressing such diverse behavioral knowledge into fixed model parameters inevitably creates an information bottleneck, limiting the model's ability to dynamically leverage relevant cross-user signals, especially when local behavioral context is sparse or insufficient~\cite{llm-esr,masr}.

Drawing inspiration from the success of in-context learning (ICL) in LLMs~\cite{gpt3,coral}, we argue that GR should not rely exclusively on the individual user sequence, but should also leverage relevant information from peers with similar interests. By introducing an explicit memory mechanism, we provide the model with dynamic guidance from users with similar behavioral trajectories, effectively bridging the gap between individual sequence modeling and collective collaborative signals. However, developing such an approach is non-trivial, as it involves two primary challenges: first, the design of a retrieval method capable of generalizing across diverse users, and second, the efficient and effective integration of the memory into existing GR models.

To address these challenges, we introduce \textbf{OMEGA}, a novel c\uline{\textbf{O}}llaborative \uline{\textbf{ME}}mory augmentation framework for \uline{\textbf{G}}enerative Recommend\uline{\textbf{A}}tion. To mitigate the redundancy of raw user sequences, we propose a latent context compression mechanism. Specifically, we employ learnable query tokens to compress variable-length user sequences into fixed-length compact embeddings. This strategy significantly reduces storage overhead while preserving the semantic information essential for accurate collaborative modeling. Furthermore, to ensure the relevance of retrieved memories, we introduce a target-aware retrieval mechanism. Notably, this mechanism incorporates similarities at both the sequence and target levels to align the retrieved memories with the specific intent of the user. Finally, to effectively fuse these external signals with the local context of the user, we employ a context-aware memory integration module. Equipped with a gated cross-attention mechanism, this module dynamically filters noise and emphasizes informative collaborative patterns for downstream decoding.

In summary, the main contributions of this work are as follows:
\begin{itemize}[leftmargin=*]
    \item We propose OMEGA, a model-agnostic generative recommendation framework that incorporates global collaborative signals via a memory-augmented paradigm.
    \item We introduce a latent context compression technique that distills the behavior sequences of users into compact representations, thereby significantly enhancing storage efficiency without compromising performance.
    \item We design a target-aware retrieval mechanism and a gated cross-attention integration strategy, which collectively ensure the retrieval of high-quality collaborative memories and their adaptive fusion with user contexts.
    \item Extensive experiments on multiple benchmark datasets demonstrate that OMEGA consistently outperforms various advanced GR baselines.
\end{itemize}

%% file: sections/2-methodology.tex
\begin{figure*}[]
    \centering
    \includegraphics[width=0.93\linewidth]{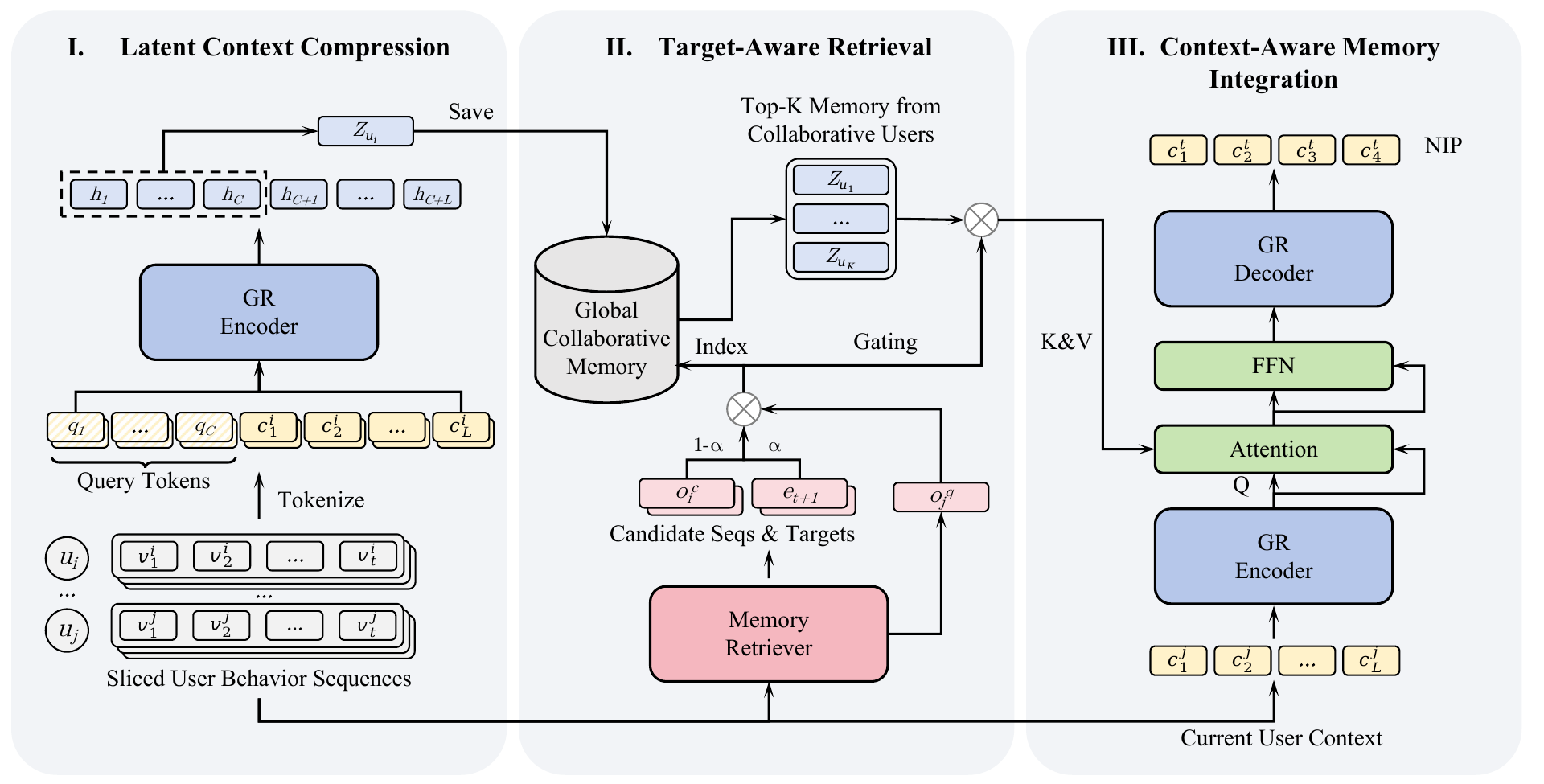}
    \captionsetup{font={small}}
    \caption{Overall architecture of OMEGA, which consists of three primary stages: (i) Latent Context Compression for distilling user behavior into compact embeddings, (ii) Target-Aware Retrieval for identifying relevant collaborative signals, and (iii) Context-Aware Memory Integration for adaptively fusing external memories with local user contexts. NIP denotes the Next Item Prediction task.}
    \label{fig:model}
\end{figure*}

\section{Methodology}
\label{sec:methodology}
In this section, we present \textbf{OMEGA}, a novel c\textbf{O}llaborative \textbf{ME}mory augmentation framework for \textbf{G}enerative recommend\textbf{A}tion as illustrated in Figure~\ref{fig:model}. 
First, we formally define the problem and introduce the backbone architecture in Section~\ref{sec:problem} and Section~\ref{sec:backbone}, respectively. 
Next, we elaborate on the construction and retrieval mechanisms of the collaborative memory bank in Section~\ref{sec:membank}.
Subsequently, Section~\ref{sec:memfusion} details the flexible memory integration strategy that fuses retrieved collaborative signals with local user contexts. 
Finally, the overall training process of the framework is discussed in Section~\ref{sec:opt}.

\subsection{Problem Formulation}
\label{sec:problem}

Let $\mathcal{U}$ and $\mathcal{V}$ denote the sets of users and items, respectively. Given a user $u \in \mathcal{U}$ and the corresponding historical interaction sequence $S_u = [v_1, v_2, \dots, v_t]$, where items are ordered chronologically by interaction timestamps, the objective of sequential recommendation is to predict the next item of interest $v_{t+1} \in \mathcal{V}$.

\paratitle{Generative Recommendation Task.}
Under the generative recommendation paradigm, the next-item prediction task is formulated as a sequence-to-sequence (Seq2Seq) generation problem. Specifically, each item is mapped to a sequence of discrete tokens that serve as its identifiers. The historical item sequence $S_u$ is consequently transformed into an input token sequence $X = [c_1, c_2, \dots, c_l]$, where $l$ denotes the sequence length. The goal is to autoregressively generate the identifier of the target item $v_{t+1}$, which is represented as a token sequence $Y = [c_{l+1}, \dots, c_{L}]$. This objective is formally defined as:
\begin{equation}
    \max_{\theta} P_{\theta}(Y \mid X) = \prod_{i=l+1}^{L} P_{\theta}(c_i \mid c_1, \dots, c_{i-1}),
\end{equation}
where $\theta$ denotes the parameters of the GR model.

\paratitle{Memory-Augmented Generative Recommendation Task.}
However, relying solely on $S_u$ is insufficient to alleviate data sparsity issues, such as the long-tail problem or short behavioral sequences of inactive users. To explicitly incorporate collaborative signals, we introduce a collaborative memory bank, denoted as $\mathcal{M}$, which stores behavioral patterns derived from the global user base. The task of memory-augmented generative recommendation is defined as follows: given the input sequence $X$ and a retrieved memory subset $m \subseteq \mathcal{M}$ containing relevant behavioral contexts, the objective is to maximize the following conditional probability:
\begin{equation}
    \max_{\theta} P_{\theta}(Y \mid X, m).
\end{equation}
In this formulation, $m$ serves as an external context to enhance the generation process. Consequently, the core research problem of this paper is twofold: 
(1) \textbf{Memory Construction and Retrieval}: How to construct a high-quality memory bank from global data and retrieve the most relevant information; 
(2) \textbf{Memory Integration}: How to effectively fuse the retrieved memory $m$ with the local sequence $X$ of the user to guide the generative process.

\subsection{Generative Backbone}
\label{sec:backbone}

Following established GR paradigms such as TIGER~\cite{tiger} and ActionPiece~\cite{actionpiece}, OMEGA utilizes a T5-like~\cite{t5} encoder-decoder architecture to model the Seq2Seq recommendation task. Notably, OMEGA serves as a model-agnostic framework compatible with various existing encoder-decoder GR models~\cite{pctx,letter}. 
Given an input token sequence $X=[c_1, \dots, c_l]$ and a target output sequence $Y=[c_{l+1}, \dots, c_{L}]$, the GR encoder first transforms the sequential context into latent representations $\bm{H} \in \mathbb{R}^{l \times d}$. Subsequently, $\bm{H}$ and $Y$ are fed into the GR decoder to extract hidden representations $\bm{\tilde{H}} \in \mathbb{R}^{(L-l) \times d}$, which are then matched against the embedding table to generate the predicted token sequence. For brevity, special tokens such as \texttt{<EOS>} and \texttt{<BOS>} are omitted. This process is formally defined as:
\begin{align}
    \bm{H} &= \text{Encoder}(X), \\
    \bm{\tilde{H}} &= \text{Decoder}(\bm{H}, Y).
\end{align}
The recommendation objective is to minimize the negative log-likelihood of the target tokens:
\begin{equation}
    \mathcal{L}_\text{GR} = -\sum_{i=1}^{L-l} \log P(Y_i \mid X, Y_{<i}),
    \label{eq:rec}
\end{equation}
where $Y_i$ represents the $i$-th target token, and $Y_{<i}$ denotes the tokens preceding $Y_i$.

\subsection{Collaborative Memory Bank}
\label{sec:membank}

This work leverages global user behaviors as external contexts to enhance recommendation performance. However, directly incorporating raw behavioral sequences into the model input incurs prohibitive computational overhead and restricts the scale of retrieved memories due to stringent latency requirements. To mitigate these issues, we propose a latent compression method that encodes each user behavior sequence into a compact, fixed-length representation to be stored within the memory bank.

\subsubsection{Fine-grained Sequence Segmentation}
In sequential recommendation, historical behavior sequences are intrinsically linked to target items. Subsequences within a complete interaction history can provide diverse collaborative signals. To preserve these fine-grained behavioral patterns, we employ sequence segmentation to partition each historical sequence into a series of prefix-target pairs. For instance, given a historical sequence $[v_1, v_2, v_3, v_4]$, the resulting segmented pairs are $\langle[v_1], v_2\rangle$, $\langle[v_1, v_2], v_3\rangle$, and $\langle[v_1, v_2, v_3], v_4\rangle$.

\subsubsection{Latent Context Compression}

As discussed in Section~\ref{sec:backbone}, the latent representation $\bm{H} \in \mathbb{R}^{l \times d}$ encapsulates the essential contextual information required for the decoder to generate recommendations. However, directly storing $\bm{H}$ in the memory bank requires massive storage space, particularly in industrial recommender systems.

Inspired by query-based compression techniques widely adopted in works like Q-Former~\cite{q-former} and LONGER~\cite{longer}, we propose a latent context compression method utilizing query tokens. Specifically, we introduce $C$ learnable query tokens, denoted as $\texttt{<Q\_1>}, \dots, \texttt{<Q\_C>}$, where each token is associated with a learnable embedding $\bm{q}_i \in \mathbb{R}^d$ for $i \in \{1, \dots, C\}$.
To perform compression, these $C$ tokens are prepended to the original input sequence $X$, to form an augmented sequence $X'$ of length $l + C$. This sequence is then processed by the GR encoder to produce the hidden states $\bm{H}' \in \mathbb{R}^{(l+C) \times d}$. We extract the first $C$ hidden states, $\bm{Z} = \bm{H}'[1:C, :]$, which correspond to the query tokens, as the \textbf{compact contextual representation}. These compressed embeddings $\bm{Z}$ are subsequently fed into the decoder for the recommendation task, which is optimized via the loss function defined in Eq.~\eqref{eq:rec}.

Through this mechanism, the contextual representation is effectively compressed from a dimension of $l \times d$ to $C \times d$. The hyperparameter $C$ offers a flexible trade-off between recommendation performance and storage efficiency. Empirically, we observe that even with $C=2$, OMEGA achieves performance comparable to that of the original GR models. This observation suggests that the semantic information in $\bm{H}$ is often sparse and highly compressible, further justifying the efficacy of the proposed latent compression method.

The compact representations $\bm{Z}$ derived from all segmented training sequences are aggregated to construct the collaborative memory bank, denoted as $\mathcal{M} = \{\bm{Z}_1, \dots, \bm{Z}_N\}$, where each $\bm{Z}_i \in \mathbb{R}^{C \times d}$ and $N$ denotes the total number of entries in the memory bank.

\subsubsection{Target-Aware Memory Retrieval}
\label{sec:retnet}

Once the memory bank is constructed, the primary challenge lies in efficiently retrieving relevant memories for a given input sequence. In Retrieval-Augmented Generation (RAG) systems, retrieval is typically executed via two mainstream paradigms: sparse retrieval~\cite{bm25} and dense retrieval~\cite{colbert}. Both approaches aim to identify the most pertinent knowledge from a source corpus based on the semantic similarity between the query and the candidates.

In our scenario, the input behavior sequence acts as the query to retrieve analogous patterns from other users. A straightforward approach would be to directly employ the compressed representations $\bm{Z}$ as both keys and values, performing retrieval based on their similarity. However, our experiments demonstrate that this approach is ineffective, likely due to representation collapse or low discriminability stemming from the constrained token space. To address this, we adopt a lightweight sequential recommender $\mathcal{R}$ (\eg SASRec~\cite{sasrec} and HSTU~\cite{hstu}) for retrieval.

Given an input sequence $S_u = [v_1, \dots, v_t]$, we first map each item to its corresponding embedding via an item embedding table $\bm{E} \in \mathbb{R}^{|\mathcal{V}| \times d_1}$, yielding the embedding sequence $[\bm{e}_1, \dots, \bm{e}_t]$. This sequence is then fed into $\mathcal{R}$ to generate a series of hidden representations $\bm{O} = \mathcal{R}([\bm{e}_1, \dots, \bm{e}_t]) = [\bm{o}_1, \dots, \bm{o}_t]$, where $\bm{o}_i \in \mathbb{R}^{d_1}$. We adopt the hidden state $\bm{o}_t$, which corresponds to the last interacted item, as the query representation of $S_u$, as it effectively summarizes the evolving preferences of the user. The retriever $\mathcal{R}$ is optimized using the standard cross-entropy loss:
\begin{equation}
    \mathcal{L}_\text{CE} = -\log \frac{\exp(g(\bm{o}_t, \bm{e}_{t+1}))}{\sum_{j \in \mathcal{V}_{\text{neg}}} \exp(g(\bm{o}_t, \bm{e}_j))},
    \label{eq:ret}
\end{equation}
where $g(\cdot)$ denotes a similarity function (e.g., dot product or cosine similarity), $t+1$ indicates the index of the ground-truth next item, and $\mathcal{V}_{\text{neg}}$ represents the set of negative candidate items.

During the retrieval phase, relying exclusively on the similarity between sequence representations is suboptimal, as it overlooks the explicit signals provided by the target item of the candidate, leading to the retrieval of irrelevant memories. To this end, we propose a \textbf{target-aware retrieval mechanism} that incorporates the similarity between the input sequence and the target item of the candidate. Formally, let $\bm{o}^q$ be the representation of the input query sequence, and $\langle\bm{o}^c_i, \bm{e}^c_i\rangle$ denote the $i$-th candidate sequence-target pair. The final retrieval score $s_i$ is defined as:
\begin{equation}
    s_i = (1-\alpha) \cdot \cos(\bm{o}^q, \bm{o}^c_i) + \alpha \cdot \cos(\bm{o}^q, \bm{e}^c_i),
\end{equation}
where $\cos(\cdot, \cdot)$ denotes cosine similarity and $\alpha$ is a hyperparameter that balances the influence of the historical sequence and the target item.

\subsection{Context-Aware Memory Integration}
\label{sec:memfusion}

Following the target-aware retrieval mechanism, we obtain the top-$K$ memories $\{\bm{Z}_i\}_{i=1}^K$ corresponding to the highest retrieval scores $\{s_i\}_{i=1}^K$ for the current input sequence $S_u$. However, for users with non-mainstream interests, the sparsity of relevant cross-user patterns may result in low-quality or noisy retrieved memories, which could potentially degrade recommendation performance. To mitigate the impact of such noise, we incorporate prior knowledge from the lightweight retriever $\mathcal{R}$ into the memories via a gating mechanism. Specifically, we convert the retrieval scores into instance-wise gating coefficients $a_i$ through a parametric transformation:
\begin{equation}
    a_i = \sigma(w \cdot s_i + b),
\end{equation}
where $w$ and $b$ are learnable parameters, and $\sigma(\cdot)$ denotes the sigmoid activation function. Intuitively, when the retriever assigns a low relevance score to a memory instance, the gate $a_i$ approaches zero, thereby effectively attenuating noisy signals. The refined memories are then represented as $\tilde{\bm{Z}}_i = a_i \cdot \bm{Z}_i$.

Subsequently, we enhance the contextual hidden state $\bm{H} \in \mathbb{R}^{l \times d}$ of the user by integrating the gated collaborative memories $\bm{M} = [\tilde{\bm{Z}}_1; \dots; \tilde{\bm{Z}}_K] \in \mathbb{R}^{(K \times C) \times d}$ via a cross-attention mechanism. Specifically, $\bm{H}$ is employed as the query, while $\bm{M}$ serves as both the key and value, enabling the model to adaptively retrieve and fuse informative signals from the external memories conditioned on the current user context. The cross-attention operation is formulated as follows:
\begin{align}
    \bm{Q} &= \bm{H}\bm{W}_Q, \bm{K} = \bm{M}\bm{W}_K, \bm{V} = \bm{M}\bm{W}_V, \\
    \hat{\bm{H}}_{\text{attn}} &= \text{Softmax}\left(\frac{\bm{Q}\bm{K}^\top}{\sqrt{d}}\right)\bm{V}, \\
    \hat{\bm{H}}_{\text{out}} &= \hat{\bm{H}}_{\text{attn}}\bm{W}_O + \bm{H},
\end{align}
where $\bm{W}_Q, \bm{W}_K, \bm{W}_V, \bm{W}_O \in \mathbb{R}^{d \times d}$ denote learnable projection matrices. 
Notably, we remove the causal masking in the cross-attention module, allowing the memory representations to attend to all positions in the original sequence, thereby facilitating more comprehensive utilization of collaborative behavioral patterns.

To further align the integrated representations with the downstream decoder, we apply a Feed-Forward Network (FFN) to transform the enhanced hidden states:
\begin{align}
    \hat{\bm{H}} &= \text{ReLU}\left(\hat{\bm{H}}_{\text{out}}\bm{W}_1+\bm{b}_1\right)\bm{W}_2+\bm{b}_2, \\
    \hat{\bm{H}} &= \hat{\bm{H}} + \hat{\bm{H}}_{\text{out}},
\end{align}
where $\bm{W}_1, \bm{W}_2 \in \mathbb{R}^{d \times d}$ and $\bm{b}_1, \bm{b}_2 \in \mathbb{R}^{d}$ are learnable parameters.
In both blocks, we apply a residual connection to preserve the semantic information of current user contexts.
Finally, the enhanced representation $\hat{\bm{H}}$ is fed into the decoder to generate the final recommendations in an autoregressive manner.

\subsection{Optimization Strategy}
\label{sec:opt}

The OMEGA framework comprises three interdependent components: the GR backbone, the latent compression model, and the lightweight retriever $\mathcal{R}$. The optimization process begins with the independent pretraining of the GR backbone and the retriever $\mathcal{R}$, which optimizes the objectives specified in Eq.~\eqref{eq:rec} and Eq.~\eqref{eq:ret}, respectively.

\subsubsection{Warm Start for Compression} 
To leverage the rich sequential knowledge captured during pretraining, we initialize the latent compression model with the parameters of the pretrained GR backbone. To maintain semantic consistency across the latent space, we freeze the vocabulary embedding table and exclusively optimize the parameters of the backbone and the learnable query tokens using the objective in Eq.~\eqref{eq:rec} until convergence. Once training is finalized, the compressed contextual representations and the corresponding retrieval representations for all candidate sequences are stored as key-value pairs within the memory bank $\mathcal{M}$.

\subsubsection{Two-Stage Finetuning}
\label{sec:two_stage}
To ensure the effective utilization of external collaborative memories within the generative process, we adopt a two-stage finetuning strategy.

\paratitle{Stage I: Attention Alignment.} We freeze the parameters of the pretrained GR backbone and exclusively optimize the cross-attention module. This stage is designed to mitigate semantic drift, thereby preventing the pretrained representations from being corrupted by the potential misalignment between the cold-start attention module and the backbone. 

\paratitle{Stage II: Joint Optimization.} Once the attention module has stabilized, we conduct end-to-end finetuning of the entire framework. This stage allows all parameters to co-adapt, further refining the synergy between the global collaborative memories and the local context of the user to achieve optimal performance.

\section{Complexity Analysis}

\paratitle{Computational Complexity.}
For memory retrieval, let the vector dimension be $d_1$ and the memory bank size be $M$.
Vector generation overhead is omitted here, as it is completed offline and introduces no latency during online inference.
We can use HNSW~\cite{hnsw} to construct the index, resulting in a search complexity of $O(d_1\cdot\log M)$.
For decoding, let the embedding and feed-forward network dimensions be $d$ and $4d$, respectively. 
The encoder and decoder each comprise $N$ layers, with input context length $L$ and the target length $4$.
Consequently, the computational complexity of context encoding is $O\left(N(4Ld^2+L^2 d)\right)$. 
The complexity of target decoding is $O(N(4^2d+4(Ld+4d^2)))$. 
The total complexity of the backbone is $O(N(16+4L+L^2)d+N(4L+16)d^2)$.
For OMEGA, the additional complexity arises from the memory integration process, which includes a cross-attention layer and an FFN layer.
Given memory length $K$, this overhead is $O(KLd+4Ld^2)$.
As the augmented and original context lengths are identical, the decoding complexity remains unchanged.
Therefore, the final complexity is $O(N(16+4L+L^2)d+N(4L+16)d^2+(KLd+4Ld^2))$.
This demonstrates that OMEGA preserves the same order of magnitude as the original backbone, ensuring practical efficiency.
Detailed experimental results are provided in Section~\ref{apx:effi}.

\paratitle{Spatial Complexity.}
The additional spatial complexity associated with OMEGA stems primarily from the maintenance of the memory bank. This overhead remains manageable, as the storage size can be dynamically adjusted to satisfy specific hardware constraints. For instance, a Least Recently Used (LRU) replacement policy can be employed to update the memory bank, thereby ensuring efficient resource utilization. Furthermore, as illustrated in Figure~\ref{fig:mr}, empirical observations demonstrate that the performance of OMEGA does not require an excessively large memory bank. Notably, utilizing only 1\% of the total capacity is sufficient to maintain robust performance.

%% file: sections/3-experiments.tex
\section{Experiments}
\label{sec:experiments}
This section evaluates the performance of the proposed OMEGA framework through comprehensive experiments and a detailed analysis.

\subsection{Experiment Setup}

\subsubsection{Dataset}
Experiments are conducted on three subsets of the Amazon 2023 review dataset~\cite{amazon2023}, specifically \emph{Musical Instruments}, \emph{Video Games}, and \emph{Industrial Scientific}. To maintain evaluation consistency, the preprocessing protocol follows the 5-core filtering approach established in previous works~\cite{s3rec,fmlp-rec}, which excludes inactive users and unpopular items with fewer than five interactions. Subsequently, user behavior sequences are organized chronologically, with the maximum sequence length truncated to 20 items. Table~\ref{tab:data_statistics} provides the statistics of the datasets after preprocessing.

\subsubsection{Baseline Models}
To evaluate the performance of OMEGA, we compare it against a diverse set of recommendation models, which can be categorized into the following two groups:
\noindent (1) \emph{SR models}:
{\textbf{SASRec}}~\cite{sasrec},
{\textbf{GRU4Rec}}~\cite{gru4rec},
{\textbf{BERT4Rec}}~\cite{bert4rec},
{\textbf{FMLP-Rec}}~\cite{fmlp-rec},
{\textbf{FDSA}}~\cite{fdsa},
{\textbf{DuoRec}}~\cite{duorec}.
\noindent (2) \emph{GR models}: 
{\textbf{TIGER}}~\cite{tiger},
{\textbf{HSTU}}~\cite{hstu},
{\textbf{LETTER}}~\cite{letter},
{\textbf{LIGER}}~\cite{liger},
{\textbf{ETEGRec}}~\cite{etegrec},
{\textbf{ActionPiece}}~\cite{actionpiece},
{\textbf{Pctx}}~\cite{pctx}.
More details of baselines are provided in Appendix~\ref{apx:baseline}

\begin{table}[]
    \centering
    \caption{Dataset Statistics. Avg.Len denotes the average length of the user action sequences.}
    \resizebox{\columnwidth}{!}{
    \renewcommand\arraystretch{0.8}
    \begin{tabular}{lrrrrr}
    \toprule
     Dataset    &\#Users   &\#Items   &\#Actions &Sparsity &Avg.Len \\
     \midrule
     Instrument &57,439  &24,587  &511,836  &99.964\% &8.91 \\
     Scientific &50,985  &25,848  &412,947  &99.969\% &8.10\\
     Game &94,762  &25,612  &814,586  &99.966\% &8.60\\
     \bottomrule
    \end{tabular}}
    \label{tab:data_statistics}
\end{table}

\begin{table*}[]
\centering
\captionsetup{font={small}}
\caption{Quantitative performance comparison of various models. Bold and underlined values indicate the best and runner-up results, respectively. ``Improv.'' represents the relative performance improvement of OMEGA over the corresponding backbone architectures. * indicates statistical significance at the $p<0.01$ level, as determined by a paired t-test. R@K and N@K are short for Recall@K and NDCG@K, respectively.}
\label{tab:main_result}
\resizebox{\textwidth}{!}{%
\renewcommand\arraystretch{0.82}
\setlength{\tabcolsep}{1.1mm}{
\begin{tabular}{lcccccccccccc}
\toprule
\multicolumn{1}{c}{\multirow{2}{*}{Methods}} & \multicolumn{4}{c}{Instrument} & \multicolumn{4}{c}{Scientific} & \multicolumn{4}{c}{Game} \\ \cmidrule(l){2-5} \cmidrule(l){6-9} \cmidrule(l){10-13}
\multicolumn{1}{r}{} & \multicolumn{1}{c}{R@5} & \multicolumn{1}{c}{R@10} & \multicolumn{1}{c}{N@5} & \multicolumn{1}{c}{N@10} & \multicolumn{1}{c}{R@5} & \multicolumn{1}{c}{R@10} & \multicolumn{1}{c}{N@5} & \multicolumn{1}{c}{N@10} & \multicolumn{1}{c}{R@5} & \multicolumn{1}{c}{R@10} & \multicolumn{1}{c}{N@5} & \multicolumn{1}{c}{N@10} \\ \midrule \midrule
GRU4Rec & 0.0324 & 0.0501 & 0.0209 & 0.0266 & 0.0202 & 0.0338 & 0.0129 & 0.0173 & 0.0499 & 0.0799 & 0.0320 & 0.0416 \\
BERT4Rec & 0.0307 & 0.0485 & 0.0195 & 0.0252 & 0.0186 & 0.0296 & 0.0119 & 0.0155 & 0.0460 & 0.0735 & 0.0298 & 0.0386 \\
SASRec & 0.0333 & 0.0523 & 0.0213 & 0.0274 & 0.0259 & 0.0412 & 0.0150 & 0.0199 & 0.0535 & 0.0847 & 0.0331 & 0.0438 \\
FMLP-Rec & 0.0339 & 0.0536 & 0.0218 & 0.0282 & 0.0269 & 0.0422 & 0.0155 & 0.0204 & 0.0528 & 0.0857 & 0.0338 & 0.0444 \\
FDSA & 0.0347 & 0.0545 & 0.0230 & 0.0293 & 0.0262 & 0.0421 & 0.0169 & 0.0213 & 0.0544 & 0.0852 & 0.0361 & 0.0448 \\
DuoRec & 0.0381 & 0.0598 & 0.0244 & 0.0314 & 0.0280 & 0.0431 & 0.0178 & 0.0226 & 0.0592 & 0.0932 & 0.0368 & 0.0477 \\ \midrule
HSTU & 0.0400 & 0.0646 & 0.0243 & 0.0322 & 0.0299 & 0.0483 & 0.0173 & 0.0232 & 0.0698 & \textbf{0.1104} & 0.0416 & 0.0547 \\
LETTER & 0.0372 & 0.0580 & 0.0246 & 0.0313 & 0.0279 & 0.0435 & 0.0182 & 0.0232 & 0.0563 & 0.0877 & 0.0372 & 0.0473 \\
LIGER & 0.0408 & 0.0630 & 0.0264 & 0.0334 & 0.0327 & {\ul 0.0528} & 0.0203 & 0.0265 & 0.0619 & 0.0969 & 0.0391 & 0.0503 \\
ETEGRec & 0.0395 & 0.0625 & 0.0257 & 0.0328 & 0.0288 & 0.0448 & 0.0189 & 0.0238 & 0.0589 & 0.0912 & 0.0385 & 0.0498 \\
ActionPiece & 0.0383 & 0.0615 & 0.0243 & 0.0318 & 0.0284 & 0.0452 & 0.0182 & 0.0236 & 0.0591 & 0.0927 & 0.0382 & 0.0490 \\ \midrule
TIGER & 0.0376 & 0.0570 & 0.0248 & 0.0310 & 0.0279 & 0.0428 & 0.0184 & 0.0232 & 0.0582 & 0.0901 & 0.0380 & 0.0482 \\
\rowcolor{blue!15}+OMEGA & {\ul 0.0437}$^*$ & 0.0647$^*$ & {\ul 0.0289}$^*$ & {\ul 0.0356}$^*$ & {\ul 0.0335}$^*$ & 0.0503$^*$ & {\ul 0.0223}$^*$ & {\ul 0.0277}$^*$ & {\ul 0.0708}$^*$ & {\ul 0.1091}$^*$ & {\ul 0.0470}$^*$ & {\ul 0.0594}$^*$ \\
Improv. & +16.22\% & +13.51\% & +16.53\% & +14.84\% & +20.07\% & +17.52\% & +21.20\% & +19.40\% & +21.65\% & +21.09\% & +23.68\% & +23.24\% \\ \midrule
Pctx & 0.0419 & {\ul 0.0655} & 0.0275 & 0.0350 & 0.0323 & 0.0504 & 0.0205 & 0.0263 & 0.0638 & 0.0981 & 0.0416 & 0.0527 \\
\rowcolor{blue!15}+OMEGA & \textbf{0.0441}$^*$ & \textbf{0.0688}$^*$ & \textbf{0.0293}$^*$ & \textbf{0.0372}$^*$ & \textbf{0.0362}$^*$ & \textbf{0.0552}$^*$ & \textbf{0.0235}$^*$ & \textbf{0.0296}$^*$ & \textbf{0.0729}$^*$ & 0.1089$^*$ & \textbf{0.0480}$^*$ & \textbf{0.0596}$^*$ \\
Improv. & +5.25\% & +5.04\% & +6.55\% & +6.29\% & +12.07\% & +9.52\% & +14.63\% & +12.55\% & +14.26\% & +11.01\% & +15.38\% & +13.09\% \\
\bottomrule
\end{tabular}}}
\end{table*}

\subsubsection{Evaluation Settings}

The performance of the sequential recommendation task is evaluated using Recall@$K$ and Normalized Discounted Cumulative Gain (NDCG)@$K$, with $K\in\{5,10\}$. Consistent with previous works~\cite{s3rec,tiger}, a \emph{leave-one-out} strategy is adopted for data partitioning. Specifically, for each user interaction sequence, the final item is reserved for testing, the penultimate item for validation, and the preceding interactions for training. To ensure a rigorous evaluation, we rank all candidate items across the complete item set.

\subsubsection{Implementation Details}

For OMEGA, we implement it using two representative GR backbones: {\textbf{TIGER}}~\cite{tiger} and {\textbf{Pctx}}~\cite{pctx}.
(a) For TIGER, we employ \texttt{sentence-t5-base}~\cite{sentence-t5} to encode textual item attributes. The RQ-VAE is configured with three codebooks of size 256, supplemented by an additional codebook for collision resolution. The hidden dimensions of the RQ-VAE encoder are hierarchically set to [2048,1024,512,256,128] and vice versa for the associated decoder. To enhance representation quality and training stability, we employ Principal Component Analysis (PCA) and Exponential Moving Average (EMA) techniques presented in previous work~\cite{mtgrec}. The backbone architecture follows a T5-like structure with a hidden size of 128, a feed-forward inner dimension of 512, and 4 attention heads (each of size 64) utilizing ReLU activation. Both the encoder and decoder consist of 4 layers.
(b) For Pctx, we strictly adhere to the original configurations and adopt the publicly released weights as pretrained weights.

In both backbones, the cross-attention module is configured with a hidden size of 128 and 2 attention heads. The number of query tokens $C$ is set to 2. We employ a lightweight HSTU as the memory retriever, consisting of 2 layers with a hidden size of 64 and 2 heads. The hyper-parameter $\alpha$ is searched within \{0.1,0.2,0.3,0.4,0.5\}. The learning rate for training the compression model is fixed at 0.001, while the learning rates for the two-stage fine-tuning process are tuned within \{0.005,0.003,0.001\} and \{0.0005,0.0003,0.0001\}, respectively. During inference, the number of retrieved memories $K$ is set to 10, and the beam size for all GR models is maintained at 50. We adopt the AdamW optimizer with a cosine learning rate scheduler for stable training.

For the baseline models, including GRU4Rec, BERT4Rec, SASRec, FMLP-Rec, FDSA, LETTER, ActionPiece, and Pctx, we report the experimental results directly from the prior study~\cite{pctx}, which are implemented based on the RecBole~\cite{recbole,recbole-new,recbole2}.

\begin{table*}[]
\centering
\captionsetup{font={small}}
\caption{Ablation study of proposed techniques on three datasets. The best and second-best results are denoted in bold and underlined fonts, respectively.}
\label{tab:ablation}
\resizebox{0.98\textwidth}{!}{%
\renewcommand\arraystretch{0.82}
\begin{tabular}{lrrrrrrrrrrrr}
\toprule
\multicolumn{1}{c}{\multirow{2}{*}{Variants}} & \multicolumn{4}{c}{Instrument} & \multicolumn{4}{c}{Scientific} & \multicolumn{4}{c}{Game} \\ \cmidrule(l){2-5} \cmidrule(l){6-9} \cmidrule(l){10-13}
 & \multicolumn{1}{c}{R@5} & \multicolumn{1}{c}{R@10} & \multicolumn{1}{c}{N@5} & \multicolumn{1}{c}{N@10} & \multicolumn{1}{c}{R@5} & \multicolumn{1}{c}{R@10} & \multicolumn{1}{c}{N@5} & \multicolumn{1}{c}{N@10} & \multicolumn{1}{c}{R@5} & \multicolumn{1}{c}{R@10} & \multicolumn{1}{c}{N@5} & \multicolumn{1}{c}{N@10} \\ \midrule \midrule

OMEGA & \textbf{0.0437} & \textbf{0.0647} & \textbf{0.0289} & \textbf{0.0356} & \textbf{0.0335} & \textbf{0.0503} & \textbf{0.0223} & \textbf{0.0277} & 
\textbf{0.0708} & \textbf{0.1091} & \textbf{0.0470} & \textbf{0.0594} \\
$\ w/o$ FT & 0.0415 & 0.0636 & 0.0277 & 0.0348 & 0.0321 & 0.0479 & 0.0213 & 0.0264 & {\ul 0.0704} & {\ul 0.1078} & {\ul 0.0465} & {\ul 0.0586} \\
$\ w/o$ Gating & {\ul 0.0429} & {\ul 0.0643} & {\ul 0.0281} & {\ul 0.0351} & {\ul 0.0330} & 0.0491 & {\ul 0.0218} & {\ul 0.0270} & 0.0695 & 0.1070 & 0.0462 & 0.0583 \\
$\ w/$ One-Stage & 0.0424 & 0.0638 & {\ul 0.0281} & 0.0350 & 0.0326 & {\ul 0.0496} & 0.0215 & {\ul 0.0270} & 0.0697 & 0.1072 & 0.0463 & 0.0584 \\
$\ w/$ Self-Mem. & 0.0407 & 0.0629 & 0.0272 & 0.0343 & 0.0317 & 0.0480 & 0.0208 & 0.0260 & 0.0650 & 0.1004 & 0.0431 & 0.0545 \\
$\ w/$ Rand-Mem. & 0.0376 & 0.0582 & 0.0250 & 0.0316 & 0.0284 & 0.0436 & 0.0189 & 0.0238 & 0.0580 & 0.0899 & 0.0379 & 0.0482 \\
Backbone & 0.0376 & 0.0570 & 0.0248 & 0.0310 & 0.0279 & 0.0428 & 0.0184 & 0.0232 & 0.0582 & 0.0901 & 0.0380 & 0.0482 \\
\bottomrule
\end{tabular}%
}
\end{table*}

\subsection{Overall Performance}

We conduct a comprehensive evaluation of OMEGA, implemented with two different backbones, against state-of-the-art ID-based sequential recommendation and GR baselines across three real-world benchmark datasets. The experimental results are summarized in Table~\ref{tab:main_result}, from which we derive the following key observations:

\begin{itemize}[leftmargin=*]
    \item GR baselines consistently outperform traditional ID-based sequential recommendation models across all evaluated datasets. This superiority can be primarily attributed to their advanced item tokenization strategies and the strong modeling capacity of generative retrieval paradigm. In particular, ActionPiece and Pctx achieve superior performance by incorporating rich contextual information into the tokenization process to produce more diverse and personalized token sequences. Moreover, HSTU demonstrates remarkable effectiveness, especially on the Games dataset, highlighting the advantages of its sequential transducer units and time-interval-aware positional encodings in modeling sequential dependencies.
    \item We employ two distinct backbones for OMEGA to comprehensively evaluate its effectiveness and generalization capability: TIGER, a representative classical GR model, and Pctx, an advanced GR approach. Experimental results show that OMEGA consistently exhibits substantial performance improvements over both backbones across all datasets, validating its robustness and strong generalization ability under different GR frameworks. Specifically, OMEGA improves TIGER by 23.68\% in terms of NDCG@5 and enhances Pctx by 15.38\% in NDCG@5. Notably, OMEGA built upon the Pctx achieves state-of-the-art performance on most evaluation metrics. These results indicate that leveraging global cross-user behavioral patterns is highly beneficial for generative recommendation. The proposed collaborative memory sharing mechanism enables the GR model to explicitly access the behaviors of similar users as reference signals prior to generating final recommendations. By combining latent context compression with context-aware memory integration, OMEGA can flexibly and efficiently exploit collaborative memories, leading to significant performance gains.
\end{itemize}

\subsection{Ablation Study}
\label{sec:ablation}

Table~\ref{tab:ablation} presents the ablation study results, evaluating the contributions of each proposed component to the overall performance. We compare OMEGA with the following five variants: 
(1) \underline{\textit{w/o} FT} excludes the joint optimization stage introduced in Section~\ref{sec:two_stage}; 
(2) \underline{\textit{w/o} Gating} removes the gating mechanism defined in Section~\ref{sec:memfusion}; 
(3) \underline{\textit{w/} One-Stage} replaces the two-stage fine-tuning with direct end-to-end fine-tuning of the entire model; 
(4) \underline{\textit{w/} Self-Mem} stores mean-pooled hidden states from the pre-trained GR encoder in the memory bank instead of the learned compressed representations; 
(5) \underline{\textit{w/} Rand-Mem} substitutes the sequential retriever with random retrieval during both training and inference.

As observed, the removal of any proposed technique leads to a performance degradation, validating the effectiveness of our design. Specifically, the two-stage optimization strategy consistently outperforms the one-stage approach, reaching a higher performance ceiling. This indicates the necessity of a warm-up phase for the fusion module to align memory signals. While using mean-pooled representations (\underline{\textit{w/} Self-Mem}) still yields improvements over the backbone, it remains inferior to our compression method, which minimizes information loss during memory encoding. Finally, random retrieval (\underline{\textit{w/} Rand-Mem}) results in a substantial performance drop, with results nearly identical to the backbone. This suggests that the model tends to ignore irrelevant memory context, highlighting that the relevance and quality of retrieved memory are critical to the efficacy of OMEGA.

\subsection{Further Analysis}

\subsubsection{Analysis of Memory Retriever and Model Ensemble}

To examine the impact of different memory retrievers, we conduct experiments with SASRec and HSTU, as shown in Table~\ref{tab:ensemble}. 
OMEGA equipped with either retriever consistently outperforms both the corresponding backbone model and the retriever alone. 
OMEGA-HSTU further achieves higher performance than OMEGA-SASRec, which is consistent with the relative performance of HSTU and SASRec. 
This observation suggests that a stronger retriever brings greater gains to OMEGA by retrieving more relevant behavioral patterns and providing richer collaborative signals for generation.
We also evaluate model ensembles to determine whether the gains of OMEGA can be attributed solely to knowledge distillation from the retriever.
Specifically, we apply z-normalization to map the predicted scores of two models to the range of 0 and 1, and then sum the normalized scores for ranking. 
The ensemble of TIGER and HSTU improves over each individual model, but still performs worse than OMEGA-HSTU. 
By contrast, the ensemble of OMEGA-HSTU and HSTU achieves higher performance than either model alone.
These results indicate that the effectiveness of OMEGA does not simply stem from mimicking the predictions of the retriever, but rather from the adaptive utilization of external contextual memory to obtain a more comprehensive representation for the current user.

\subsubsection{Impact of the Number of Compression Tokens and Memory Ratio}

\begin{figure}[]
    \centering
    \includegraphics[width=0.9\columnwidth]{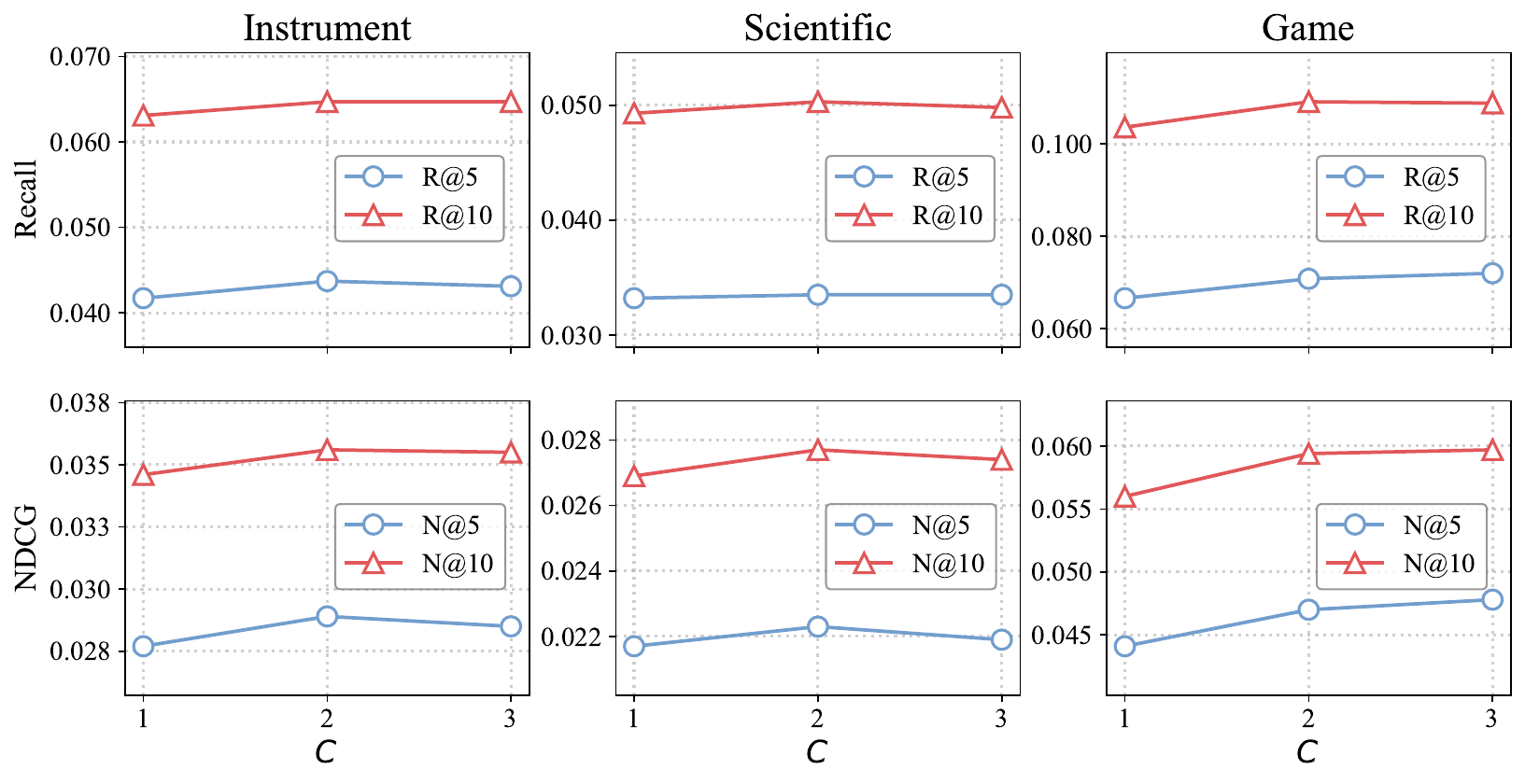}
    \captionsetup{font={small}}
    \caption{Performance \wrt the number of compression query tokens $C$.}
    \label{fig:cr}
\end{figure}

\begin{figure}[]
    \centering
    \includegraphics[width=0.9\columnwidth]{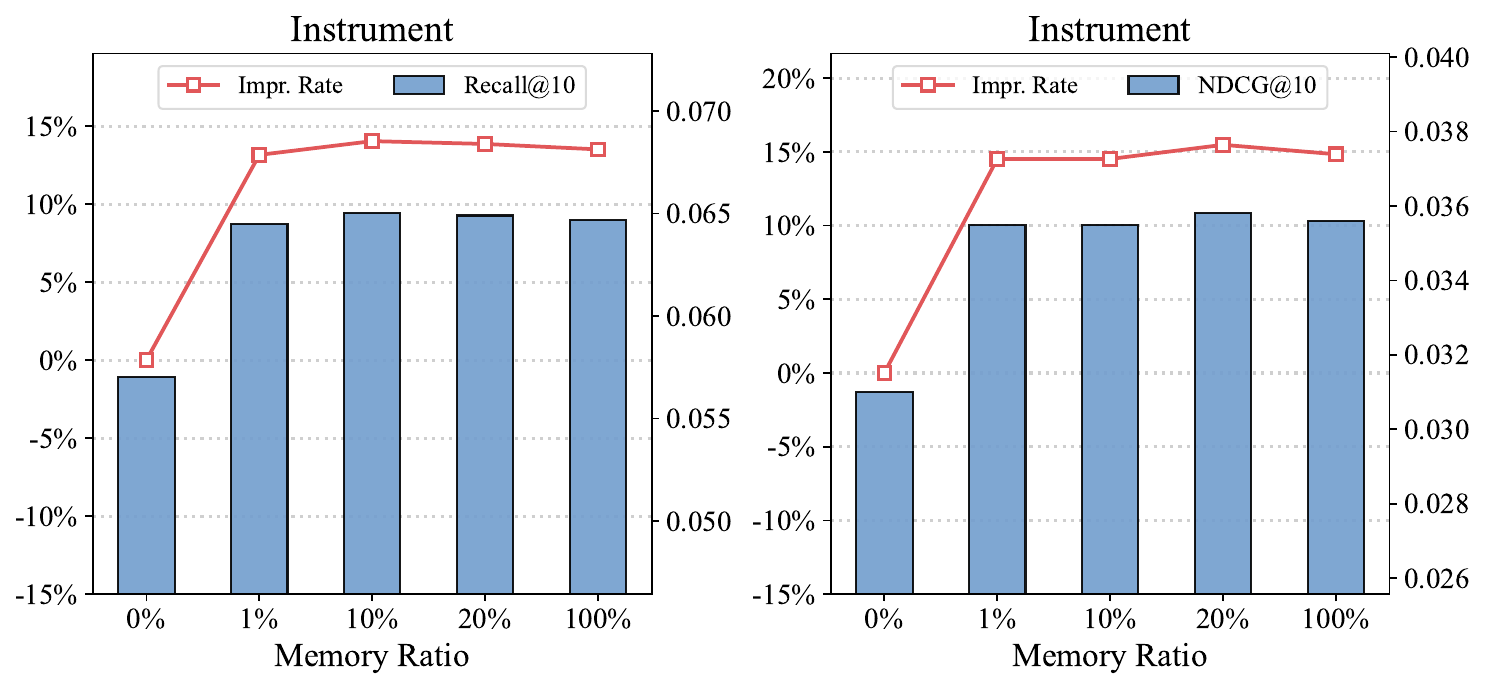}
    \captionsetup{font={small}}
    \caption{Performance \wrt memory ratio on Instrument.}
    \label{fig:mr}
\end{figure}

We study the effect of the number of compression tokens by varying it from 1 to 3, with the results reported in Figure~\ref{fig:cr}. 
Overall performance improves when increasing $C$ from 1 to 2 and remains stable from $C=2$ to $C=3$ on the Instrument and Scientific datasets, while it continues to improve from $C=1$ to $C=3$ on the Game dataset. 
These results suggest that higher-quality contextual representations generally lead to better performance, but the benefit saturates beyond a certain point. 
Specifically, two compression tokens are sufficient to preserve the essential information for Instrument and Scientific, whereas Game requires a larger number of tokens to capture more complex user behaviors.
We further analyze the impact of the memory bank size by randomly sampling a fraction $r$ of entries from the original memory bank, where $r \in \{1\%, 10\%, 20\%, 100\%\}$. 
As shown in Figure~\ref{fig:mr}, using only 1\% of the memory bank already achieves performance comparable to that obtained with the full memory bank. 
This observation indicates that OMEGA captures high-level and generalizable behavior patterns from memory, rather than relying on fine-grained user behavior signals, highlighting the storage efficiency of OMEGA.

\subsubsection{Impact of Retrieval Coefficient and Retrieval Number}

\begin{figure}[]
    \centering
    \includegraphics[width=0.9\columnwidth]{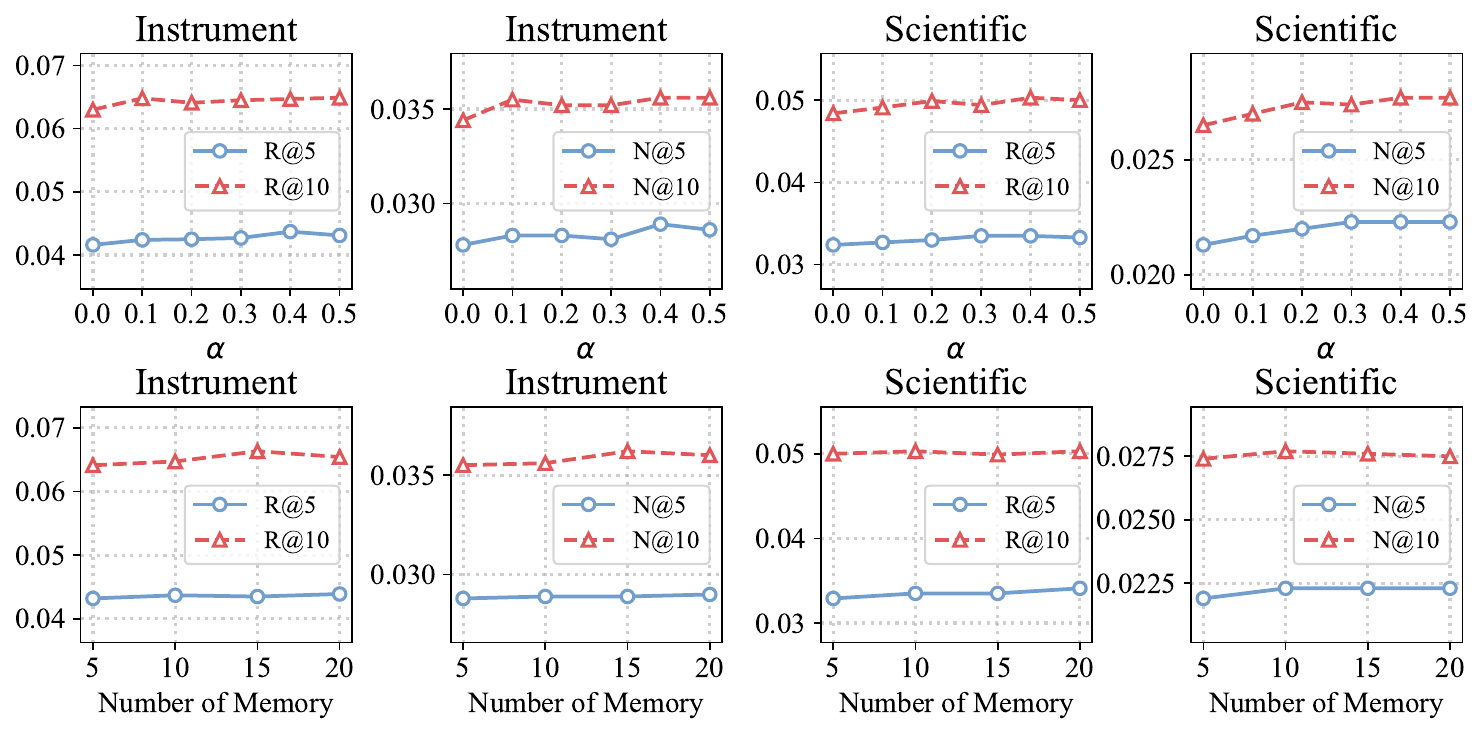}
    \captionsetup{font={small}}
    \caption{Performance \wrt the retrieval coefficient $\alpha$ and number of retrieved memory $K$.}
    \label{fig:alpha}
\end{figure}

To systematically examine the effect of the retrieval coefficient $\alpha$ on model performance, we vary $\alpha$ from 0 to 0.5 with a step size of 0.1. 
The results are shown in Figure~\ref{fig:alpha} and reveal several observations. 
When $\alpha$ is set to 0, \ie retrieval relies solely on sequence-level similarity, performance degrades across all metrics, indicating that target-level similarity provides useful signals during retrieval. 
In addition, the optimal value of $\alpha$ varies across datasets, suggesting that selecting an appropriate retrieval coefficient is important for achieving strong performance.
We further analyze the impact of the number of retrieved memory entries $K$ by varying $K$ within \{5, 10, 15, 20\}. 
As shown in Figure~\ref{fig:alpha}, performance is relatively insensitive to the choice of $K$, which indicates that a small number of high-quality memory entries is sufficient for OMEGA to extract meaningful behavioral patterns.

\begin{table}[]
\centering
\captionsetup{font={small}}
\caption{The results of different memory retrievers and model ensemble. The best and second-best results are denoted in bold and underlined fonts, respectively.}
\label{tab:ensemble}
\resizebox{0.97\columnwidth}{!}{%
\renewcommand\arraystretch{0.9}
\begin{tabular}{crrrrrr}
\toprule
\multirow{2}{*}{Variants} & \multicolumn{2}{c}{Instrument} & \multicolumn{2}{c}{Scientific} & \multicolumn{2}{c}{Game} \\
\cmidrule(l){2-3} \cmidrule(l){4-5} \cmidrule(l){6-7}
 & \multicolumn{1}{c}{R@10} & \multicolumn{1}{c}{N@10} & \multicolumn{1}{c}{R@10} & \multicolumn{1}{c}{N@10} & \multicolumn{1}{c}{R@10} & \multicolumn{1}{c}{N@10} \\ \midrule \midrule
SASRec & 0.0523 & 0.0274 & 0.0412 & 0.0199 & 0.0847 & 0.0438 \\
HSTU & 0.0646 & 0.0322 & 0.0483 & 0.0232 & {\ul 0.1104} & 0.0547 \\
TIGER & 0.0570 & 0.0310 & 0.0428 & 0.0232 & 0.0901 & 0.0482 \\
OMEGA-SASRec & 0.0623 & 0.0341 & 0.0483 & 0.0266 & 0.0950 & 0.0510 \\
OMEGA-HSTU & {\ul 0.0647} & {\ul 0.0356} & {\ul 0.0503} & {\ul 0.0277} & 0.1091 & {\ul 0.0594} \\ \midrule
\multicolumn{7}{c}{Model Ensemble} \\ \midrule
TIGER+HSTU & 0.0622 & 0.0343 & 0.0473 & 0.0260 & 0.0978 & 0.0529 \\
OMEGA+HSTU & \textbf{0.0684} & \textbf{0.0375} & \textbf{0.0528} & \textbf{0.0293} & \textbf{0.1146} & \textbf{0.0623} \\
\bottomrule
\end{tabular}%
}
\end{table}

\subsubsection{Inference Efficiency Analysis}
\label{apx:effi}

To evaluate the efficiency of OMEGA, the number of trainable parameters, total training time, memory retrieval overhead and inference speed are compared against the backbone model in Table~\ref{tab:efficiency}. Regarding the parameter scale, OMEGA introduces approximately 7\% additional trainable parameters while delivering substantial performance improvements. In terms of total training time, OMEGA incurs higher overhead than the backbone due to the fine-tuning stage. However, this fine-tuning process is significantly more efficient than pre-training, requiring only 30.0\% of the pre-training time for the Instrument dataset, 60.0\% for the Scientific dataset, and 13.2\% for the Game dataset, which remains computationally feasible in practice. To assess inference speed, the inference time on the test set is measured using 4 NVIDIA RTX3090 GPUs with a batch size of 128 per device. The results indicate that the inference speed of OMEGA is approximately 1\% slower than that of the backbone. This marginal difference demonstrates that OMEGA introduces minimal computational bottlenecks, thereby preserving the efficiency of the original backbone architecture. Furthermore, the cost of memory retrieval is negligible compared to inference, accounting for less than 2\% of the inference time. This finding indicates that the memory retrieval stage does not constitute an efficiency bottleneck.

\begin{table}[H]
\centering
\captionsetup{font={small}}
\caption{Efficiency comparison between OMEGA and the backbone regarding number of trainable parameters, total training time, memory retrieval time and inference time.}
\label{tab:efficiency}
\resizebox{0.98\columnwidth}{!}{%
\renewcommand\arraystretch{1.1}
\setlength{\tabcolsep}{0.5mm}{
\begin{tabular}{lrrrrrrrrrr}
\toprule
\multicolumn{1}{c}{\multirow{2}{*}{Model}} & \multirow{2}{*}{\#Params} & \multicolumn{3}{c}{Instrument} & \multicolumn{3}{c}{Scientific} & \multicolumn{3}{c}{Game} \\ \cmidrule(l){3-5} \cmidrule(l){6-8} \cmidrule(l){9-11}
\multicolumn{1}{c}{} &  & \#Train & \#Ret. &\#Infer. & \#Train & \#Ret. & \#Infer. & \#Train & \#Ret. & \#Infer. \\ \midrule \midrule
TIGER & 2.76M & 1.5h & 0s & 16.04s & 1.0h & 0s & 13.50s & 8.3h & 0s & 25.52s \\
OMEGA & 2.95M & 2.0h & 0.12s & 16.18s & 1.6h & 0.08s & 13.65s & 9.4h & 0.45s & 25.89s \\
\bottomrule
\end{tabular}%
}}
\end{table}

\subsubsection{Storage Efficiency Analysis}

OMEGA introduces additional storage overhead due to the maintenance of a memory bank and retrieval-related objects. 
In our offline experiments, we precompute and store the retrieval indices and scores for each sample. For the Instrument dataset, this requires 104.4 MB of storage, while the compressed representations of all training samples (approximately 340k samples in float32 format) occupy 332 MB.
For industrial deployment, OMEGA does not necessarily require storing every individual item in the memory bank. As shown in Figure~\ref{fig:mr}, OMEGA is not strongly dependent on a large memory bank to achieve competitive performance. Therefore, memory reduction techniques, such as clustering-based pruning, can be applied to compress the memory bank and corresponding retrieval vectors, thereby reducing search complexity while maintaining scalability and efficiency in large-scale settings.

\subsubsection{Generalization Across Scenarios}

To further evaluate the effectiveness of OMEGA in scenarios involving longer interaction sequences and larger user bases, we conduct additional experiments on two datasets: MovieLens-1M and Baby from Amazon Review 2023.

\paratitle{Data Preprocessing.}
For MovieLens-1M (ML-1M), we remove interactions with ratings lower than 4 and apply 5-core filtering to exclude inactive users and infrequent items. For Baby, we follow the same preprocessing strategy described in Section~\ref{sec:experiments}. The maximum sequence length is set to 100 for ML-1M and 20 for Baby, respectively. Statistics of the preprocessed datasets are summarized in Table~\ref{tab:addi_data}.

\begin{table}[]
\centering
\caption{Dataset Statistics. Avg.Len denotes the average length of the user action sequences.}
\resizebox{\columnwidth}{!}{
\renewcommand\arraystretch{0.8}
\begin{tabular}{lrrrrr}
\toprule
 Dataset    &\#Users   &\#Items   &\#Actions &Sparsity &Avg.Len \\
 \midrule
 ML-1M &6,034  &3,125  &574,376  &96.954\% &95.19 \\
 Baby &150,777  &36,013  &1,241,083  &99.977\% &8.23\\
 \bottomrule
\end{tabular}}
\label{tab:addi_data}
\end{table}

\paratitle{Overall Performance.}
As shown in Table~\ref{tab:res_addi}, OMEGA consistently improves recommendation performance on both ML-1M and Baby. On ML-1M, a long-sequence setting, OMEGA improves Recall@10 by 10.7\% over TIGER, suggesting its effectiveness in modeling long-range dependencies. On Baby, which involves a larger user base, OMEGA further achieves a 20.2\% improvement in Recall@10, demonstrating its robustness in larger-scale recommendation scenarios.

\begin{table}[]
\centering
\captionsetup{font={small}}
\caption{Performance comparison on ML-1M and Baby. Bold and underlined values indicate the best and second-best results, respectively.}
\resizebox{\columnwidth}{!}{
\renewcommand\arraystretch{0.7}
\setlength{\tabcolsep}{1mm}{
\begin{tabular}{lcccc}
\toprule
\multirow{2}{*}{Model} & \multicolumn{2}{c}{ML-1M} & \multicolumn{2}{c}{Baby} \\ \cmidrule(l){2-3} \cmidrule(l){4-5}
   & Recall@10    & NDCG@10   & Recall@10   & NDCG@10   \\ \midrule \midrule
SASRec           & 0.2267           & 0.1214          & 0.0352          & 0.0178          \\
FMLPRec          & 0.2257           & 0.1197          & 0.0353          & 0.0186          \\
HSTU             & \underline{0.2416}           & \underline{0.1308}          & \underline{0.0465}          & \underline{0.0238}          \\ \midrule
TIGER            & 0.2348           & 0.1307          & 0.0410          & 0.0225          \\
+OMEGA           & \textbf{0.2599}  & \textbf{0.1463} & \textbf{0.0493} & 
\textbf{0.0268} \\
\bottomrule
\end{tabular}}}
\label{tab:res_addi}
\end{table}

\subsubsection{Generalization Across Backbone Methods}

To further evaluate the generalization ability of OMEGA across diverse backbone methods, we apply OMEGA to SETRec~\cite{setrec}, an identifier order-agnostic generative recommendation method. Following the original setting, we adopt the official implementation with T5-small as the LLM backbone. The results are reported in Table~\ref{tab:res_setrec}.
OMEGA consistently improves the performance of SETRec on both Instrument and Scientific, further demonstrating its compatibility with different backbones. These results suggest that explicit collaborative memory enhancement has the potential to complement a broader range of generative recommendation methods.

\begin{table}[]
\centering
\captionsetup{font={small}}
\caption{Performance comparison with SETRec as the backbone.}
\resizebox{\columnwidth}{!}{
\renewcommand\arraystretch{0.7}
\setlength{\tabcolsep}{1mm}{
\begin{tabular}{lcccc}
\toprule
\multirow{2}{*}{Model} & \multicolumn{2}{c}{Instrument} & \multicolumn{2}{c}{Scientific} \\ \cmidrule(l){2-3} \cmidrule(l){4-5}
   & Recall@10    & NDCG@10   & Recall@10   & NDCG@10   \\ \midrule \midrule
SETRec & 0.0620 & 0.0324 & 0.0453 & 0.0234 \\
+OMEGA & \textbf{0.0664} & \textbf{0.0353} & \textbf{0.0505} & \textbf{0.0259} \\
\bottomrule
\end{tabular}}}
\label{tab:res_setrec}
\end{table}

%% file: sections/4-related-work.tex
\section{Related Work}
\label{sec:related}

\subsection{Generative Recommendation}
\label{sec:rel_gr}
Traditional SR approaches rely on substantial item embedding tables.
However, this open-set setting faces scalability challenges~\cite{mwuf,trm} as item catalogs expand.
To overcome these limitations, the GR paradigm introduces item tokenization. By representing items as semantic IDs within a closed vocabulary, similar to LLMs, GR reformulates recommendation as a Seq2Seq generation task.
Existing research has explored diverse item tokenization strategies, encompassing approaches based on item metadata~\cite{idgenrec,gpt4rec,grace}, hierarchical clustering~\cite{howtoindex,seater,eager}, and Vector Quantization (VQ)~\cite{tiger,rpg,mbgen}.
Within the VQ-based approaches, early methods such as TIGER and LC-Rec utilize RQ-VAE. Subsequent works have investigated alternative quantization techniques to improve ID quality, such as Finite Scalar Quantization (FSQ) in RecGPT~\cite{recgpt}, RQ-KMeans in OneRec~\cite{onerec}, and Optimized Product Quantization (OPQ) in RPG~\cite{rpg}.
Recent advancements have enhanced GR across several dimensions. Specifically, LETTER~\cite{letter}, MMQ~\cite{mmq} and MMQ-v2~\cite{mmq-v2} integrate collaborative signals into the item tokenization, while hybrid frameworks such as LIGER~\cite{letter} and COBRA~\cite{cobra} bridge the gap between dense and generative retrieval. Beyond that, ETEGRec~\cite{etegrec} and BLOGER~\cite{bloger} propose a unified paradigm that jointly optimizes tokenization and recommendation within a single stage. 
CoFiRec~\cite{cofirec} and GRACE~\cite{grace} highlight the significance of explicit coarse-to-fine semantic IDs to maintain structural hierarchy.
Furthermore, ActionPiece~\cite{actionpiece} and Pctx~\cite{pctx} introduce dynamic contextual tokenization to adaptively incorporate context information into tokenization process for better behavior modeling.
In contrast to these approaches, OMEGA explores an orthogonal direction in GR by leveraging memory mechanisms to explicitly perceive similar users' interests during target user modeling, thereby improving preference representation.

\subsection{Memory-Augmented Recommendation}
Recently, a growing body of research in recommender systems has explored memory-enhanced architectures beyond purely parametric models.
Prior studies such as CoRAL~\cite{coral} retrieve historical user–item interactions to better align LLM-based reasoning with recommendation-specific knowledge.
RaSeRec~\cite{raserec} introduces a dynamic memory bank to mitigate preference drift and long-tail forgetting in sequential recommendation.
To address ultra-long user sequences in industrial systems, several studies have investigated memory mechanisms for more efficient behavior modeling.
LMN~\cite{lmn} compresses user behaviors into compact memory blocks to capture shared interests among similar users while preserving long-term preferences.
MARM~\cite{marm} reduces attention complexity from $O(n^2)$ to $O(n)$ through memory augmentation, enabling scalable modeling of long sequences.
VISTA~\cite{vista} summarizes ultra-long user histories into cached virtual tokens for efficient handling of extended interaction sequences, and LEMUR~\cite{lemur} incrementally accumulates multimodal representations via a memory bank to support efficient end-to-end training at scale.
Building on these advances, we propose OMEGA, which incorporates explicit collaborative memory into the GR models.
By leveraging latent context compression and target-aware retrieval, OMEGA explicitly utilizes cross-user collaborative signals to enhance recommendation performance.

%% file: sections/5-conclusion.tex
\section{Conclusion}
\label{sec:conclusion}

In this paper, we propose \textbf{OMEGA}, a collaborative memory augmentation framework for GR. By leveraging user behavior sequences as a corpus-level knowledge source, OMEGA constructs a collaborative memory bank to provide external contextual guidance for more accurate recommendations. To support efficient memory utilization, OMEGA incorporates latent context compression and lightweight target-aware retrieval with gated cross-attention. Extensive experiments on real-world datasets demonstrate that OMEGA is compatible with existing GR approaches and consistently improves their performance.
In future work, we plan to enrich the memory with more diverse user behaviors, such as search history, and explore the scalability of OMEGA with larger foundation models.

%% file: sections/6-appendix.tex
\section{Appendix}

\subsection{Baselines}
\label{apx:baseline}
(1) \emph{SR models}: 
\begin{itemize}
    \item {\textbf{GRU4Rec}}~\cite{gru4rec} employs Gated Recurrent Units (GRUs) to effectively capture evolving sequential patterns within user-item interactions.
    \item {\textbf{BERT4Rec}}~\cite{bert4rec} utilizes a bidirectional self-attention mechanism coupled with a Cloze task objective to model complex item dependencies.
    \item {\textbf{SASRec}}~\cite{sasrec} leverages a unidirectional Transformer-based architecture to predict the next item of interest by modeling long-term user sequences. 
    \item {\textbf{FMLP-Rec}}~\cite{fmlp-rec} introduces an all-MLP framework equipped with learnable frequency filters to mitigate noise in sequential modeling. 
    \item {\textbf{FDSA}}~\cite{fdsa} integrates item-level and feature-level correlations through a dual-stream self-attentive network to enhance representation learning. 
    \item {\textbf{DuoRec}}~\cite{duorec} proposes a contrastive learning framework with model-level data augmentation to address the issue of representation degeneration.
\end{itemize}

(2) \emph{GR models}: 
\begin{itemize}
    \item {\textbf{TIGER}}~\cite{tiger} adopts the RQ-VAE quantization technique to map rich item attributes into discrete semantic IDs for autoregressive generation. 
    \item {\textbf{HSTU}}~\cite{hstu} introduces a hierarchical sequential transduction architecture tailored for generative recommendation, prioritizing high efficiency and superior scaling performance. 
    \item {\textbf{LETTER}}~\cite{letter} enhances item tokenization by incorporating collaborative signals into the RQ-VAE~\cite{rqvae} to better capture user-item affinities.
    \item {\textbf{LIGER}}~\cite{liger} is a hybrid framework combining generative retrieval with dense retrieval for sequential recommendation.
    \item {\textbf{ETEGRec}}~\cite{etegrec} bridges the item tokenization with generative recommendation through mutual alignment objectives.
    \item {\textbf{ActionPiece}}~\cite{actionpiece} presents a dynamic, context-aware tokenization strategy that adaptively constructs item identifiers based on diverse feature sets. 
    \item {\textbf{Pctx}}~\cite{pctx} utilizes a personalized context-aware tokenization method that leverages latent representations from DuoRec to generate user-specific semantic IDs.
\end{itemize}

\subsection{Impact of Different User Groups}
\begin{figure}[H]
    \centering
    \includegraphics[width=\columnwidth]{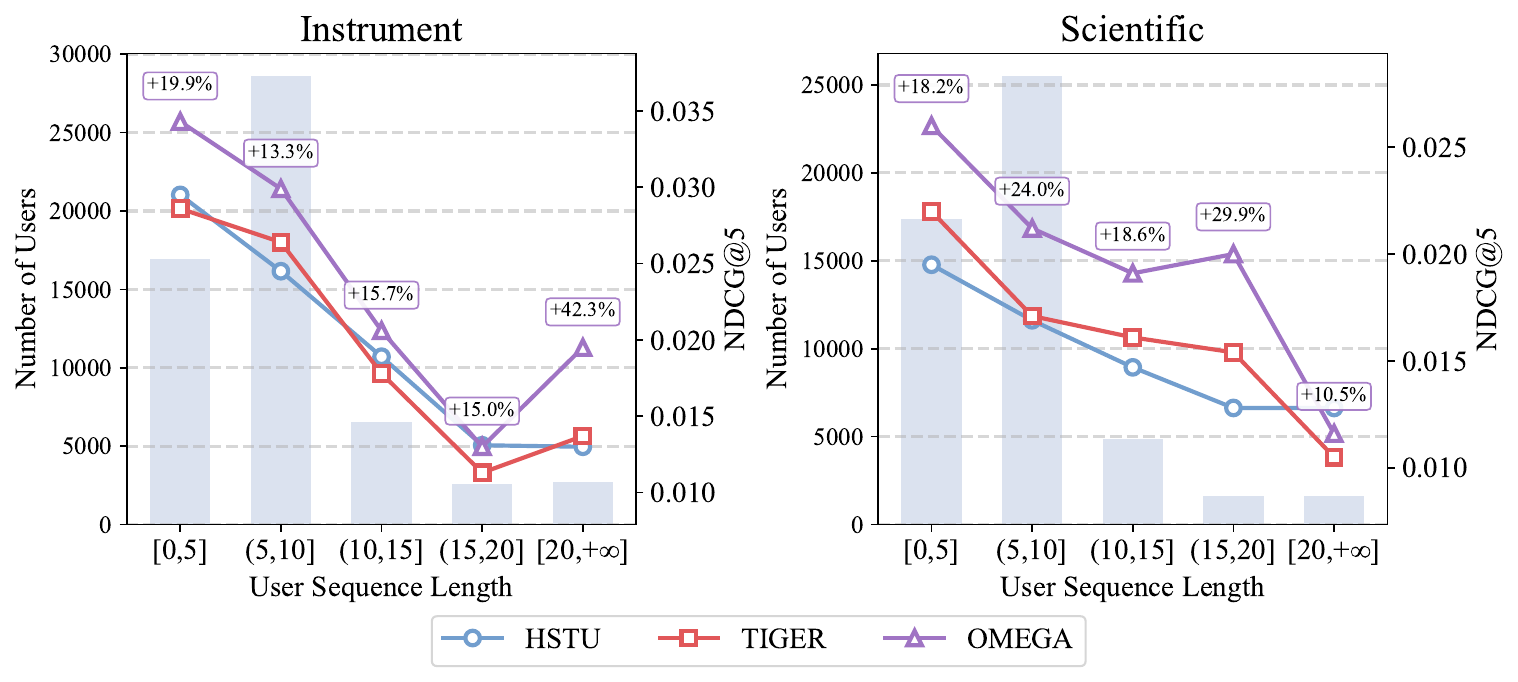}
    \captionsetup{font={small}}
    \caption{Performance \wrt the users with different levels of interaction sparsity. The relative improvement of OMEGA over the TIGER baseline is annotated within the figure.}
    \label{fig:user_group}
\end{figure}

To assess the influence of OMEGA across cohorts with varying activity levels, users in the test set are categorized into distinct groups based on the frequency of their interactions within the training set. As illustrated by the results in Figure~\ref{fig:user_group}, the majority of users in the Instrument and Scientific datasets exhibit fewer than 10 interactions. OMEGA consistently demonstrates superior performance across all user groups for both datasets. Notably, OMEGA yields a substantial improvement of 42.3\% for the user group with more than 20 interactions in the Instrument dataset. In contrast, for the Scientific dataset, the performance gains of OMEGA are relatively more pronounced among user groups with fewer than 20 interactions.
These results validate the effectiveness of OMEGA, as it enhances the preference modeling capability of the GR backbone through the introduction of diverse collaborative contexts.

\subsection{Impact of Different Item Groups}

\begin{figure}[]
    \centering
    \includegraphics[width=\columnwidth]{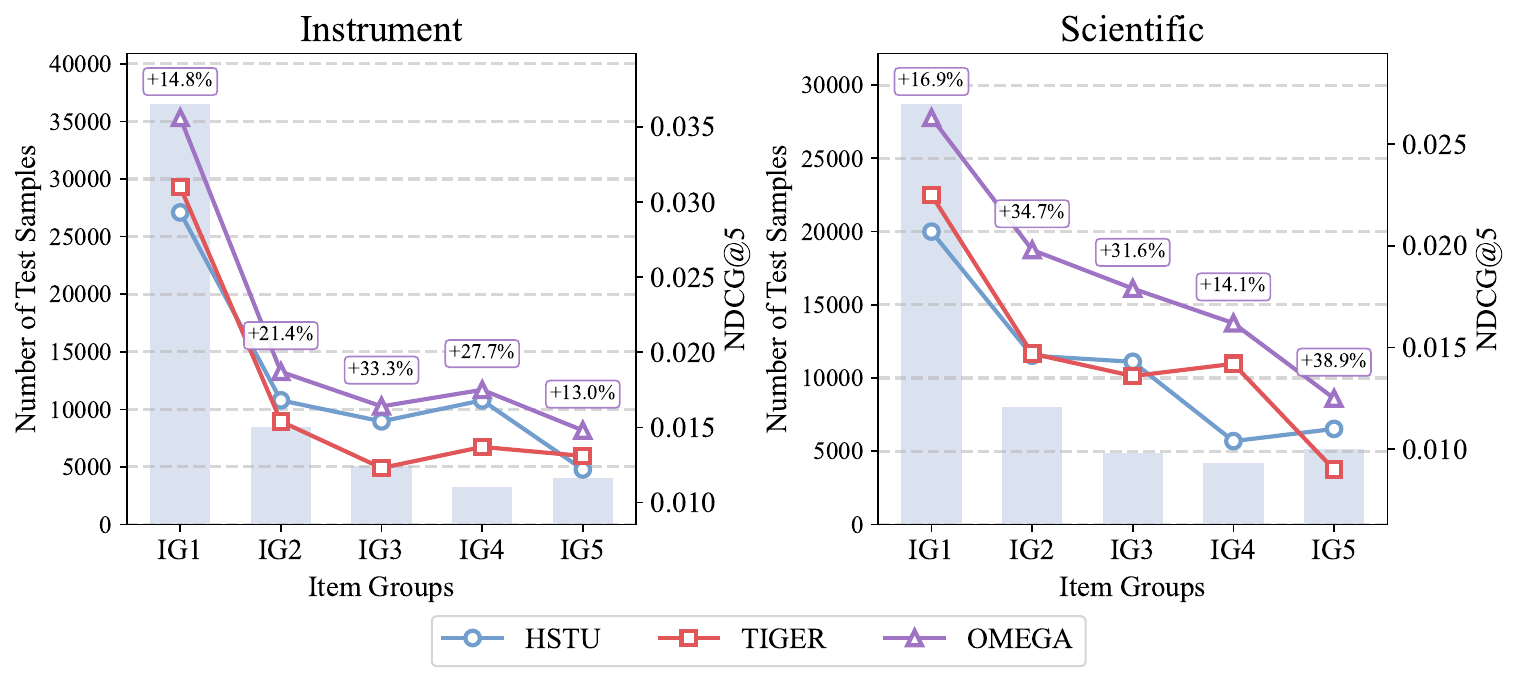}
    \captionsetup{font={small}}
    \caption{Performance \wrt the items with varying levels of interaction sparsity. The relative improvement of OMEGA over the TIGER baseline is annotated within the figure.}
    \label{fig:item_group}
\end{figure}

Figure~\ref{fig:item_group} illustrates the performance of HSTU, TIGER, and OMEGA across five item groups characterized by varying levels of interaction sparsity on the Instrument and Scientific datasets. Specifically, items are partitioned into five groups of comparable size based on the frequency of their interactions. From \texttt{IG1} to \texttt{IG5}, the interaction frequency is in descending order. A long-tail distribution of the test sample size is observed from \texttt{IG1} to \texttt{IG5} across both datasets. Furthermore, the results indicate that less popular items derive greater benefits from OMEGA, as evidenced by the performance in \texttt{IG2}, \texttt{IG3}, and \texttt{IG4} on the Instrument, and \texttt{IG2}, \texttt{IG3}, and \texttt{IG5} on the Scientific. This trend likely arises because these items are insufficiently modeled by the backbone due to the relative scarcity of training data compared to popular items. By incorporating external informative memory, the GR model accesses enriched information, thereby facilitating more accurate predictions.

